\documentclass{article}
\usepackage[T1]{fontenc}
\usepackage[utf8]{inputenc}
\usepackage{ismir}
\usepackage{amsmath,amssymb,cite}
\usepackage{graphicx}
\usepackage{xurl}
\usepackage[bookmarks=false,hidelinks]{hyperref}
\usepackage{color}
\usepackage{siunitx}
\title{DDSynth-RL: Audio Synthesizer Inversion \\via Discrete Diffusion with Reinforcement Learning}

\multauthor
  {Tristan Wu$^1$ \hspace{0.35cm} Daniel Chin$^2$ \hspace{0.35cm} Junan Zhang$^3$ \hspace{0.35cm} Junyan Jiang$^2$ \hspace{0.35cm} Yansen Jing$^4$ \hspace{0.35cm} Gus Xia$^{2,5}$}
  {$^1$ Computational Media and Art, The Hong Kong University of Science and Technology (Guangzhou)\\
   $^2$ New York University Shanghai \quad $^3$ The Chinese University of Hong Kong, Shenzhen\\
   $^4$ Department of Automation, Tsinghua University\\
   $^5$ Mohamed bin Zayed University of Artificial Intelligence\\
   {\tt\small wwu252@connect.hkust-gz.edu.cn, daniel.chin@nyu.edu, junanzhang@link.cuhk.edu.cn}\\
   {\tt\small jj2731@nyu.edu, jys23@mails.tsinghua.edu.cn, gus.xia@mbzuai.ac.ae}
  }

\def\authorname{T. Wu, D. Chin, J. Zhang, J. Jiang, Y. Jing, and G. Xia}

\begin{document}

\maketitle

\begin{abstract}
\looseness=-1
Synthesizer inversion is challenging for two main reasons: 1) Distinct parameter configurations can produce perceptually similar sounds. 2) Parameter-space losses often fail to reflect rendered audio similarity, while the synthesizer being a non-differentiable black box prevents simple audio-domain supervision. To address the one-to-many mapping induced by the first challenge, we formulate synthesizer inversion as conditional generation over discrete synthesizer parameters and use masked discrete diffusion as the generator. This treatment additionally avoids the fixed-order assumption of autoregressive models and the continuous-relaxation mismatch of flow matching when modeling categorical synthesizer controls. To address the second challenge, we further fine-tune the model with GRPO-style audio-domain rewards computed from rendered outputs. Experiments on Dexed show that, after supervised training, the discrete diffusion model is competitive with autoregressive and flow-matching baselines, and reward-based fine-tuning further improves out-of-domain audio matching performance. Code and demos are available at: \url{https://github.com/DDSynth-RL/DDSynthRL}.
\end{abstract}

\section{Introduction}\label{sec:introduction}

\begin{figure}[t]
  \centering
  \includegraphics[alt={Comparison of three generative modeling frameworks for synthesizer inversion},width=\linewidth]{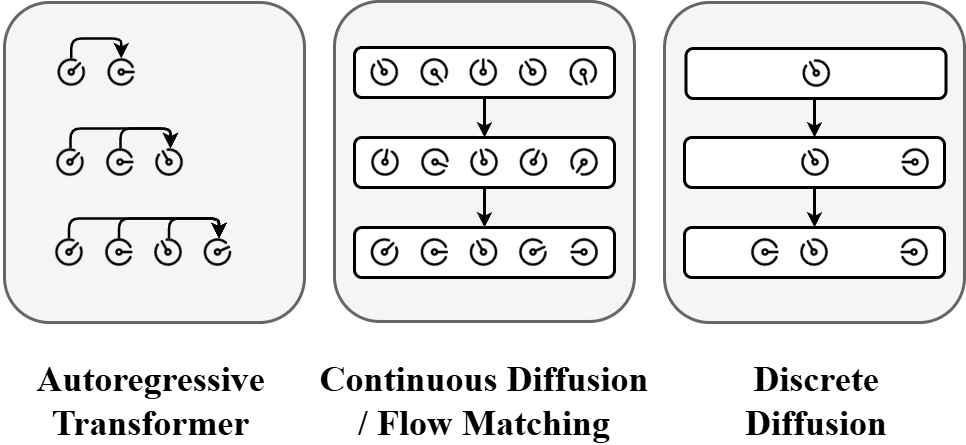}
  \caption{Comparison of three generative modeling frameworks for synthesizer inversion.}
  \label{fig:framework_comparison}
\end{figure}

Sound synthesizers underpin a wide range of creative practices including music production, film and game sound design, and new media art. Their controls are closely tied to human auditory perception and creative practices.

In music AI research, symbolic music modeling has long been a central topic, but ``symbolic'' usually refers to notes and scores. Prior work has studied audio-to-symbol extraction\cite{lin2021unified,gardner2021mt3}, text-to-symbolic-music generation\cite{yuan2024chatmusician,wang2025notagen,lu2023musecoco}, and real-time symbolic music generation and accompaniment\cite{xia2015spectral,wu2025adaptive}, using representations such as MIDI, MusicXML, and ABC notation. However, note-level musical symbols such as MIDI are not sufficient to describe all information contained in music audio: the same MIDI notes can lead to very different sounds depending on the synthesizer parameters and sound-design choices used to render them.

We therefore view synthesizer parameters as an important but underexplored symbolic channel for music AI research. The corresponding audio-to-symbol task is synthesizer inversion: inferring synthesizer parameters from audio\cite{horner1993machine,garcia2001growing,mitchell2005frequency,heise2009automated,esling2019flow,le2021improving,chen2022sound2synth,hayes2025audio,masuda2023improving,uzrad2024diffmoog,yang2023white}. Successful synthesizer inversion allows sound designers to inspect how a sample might be recreated and explore nearby sounds through parameter editing.

Synthesizer inversion is challenging for two main reasons. First, it is a one-to-many conditional generation problem: multiple parameter configurations may produce perceptually similar timbres\cite{hayes2025audio}. Second, parameter-space losses do not necessarily reflect perceptual similarity between rendered sounds\cite{shin2025synthrl}.

The first challenge makes deterministic regression insufficient and motivates a generative formulation. However, common generative models have limitations for synthesizer parameters. Autoregressive models require a fixed generation order, which can impose an artificial bias on parameters that are not naturally sequential. Continuous diffusion and flow matching models avoid this ordering issue, but they model categorical controls through continuous representations, introducing a mismatch between the model space and the discrete parameter space. To address this challenge, we introduce a \textbf{discrete diffusion} model that performs noise injection and denoising directly in parameter-token space. At each step, the model predicts all masked parameters simultaneously and iteratively decodes high-confidence tokens. Figure 1 summarizes the contrast between autoregressive generation, continuous diffusion and flow matching, and our masked discrete diffusion formulation.

The second challenge motivates audio-space supervision. A direct approach would be to render predicted parameters and compare the rendered audio with the target. In most practical settings, however, the synthesizer is treated as a \textit{black box}: the mapping from parameters to audio is not differentiable. Audio-space losses therefore cannot be directly backpropagated through the renderer, motivating reward-based optimization from rendered outputs. To address this challenge, we further fine-tune the model with Group Relative Policy Optimization (\textbf{GRPO}), using rewards computed from audio metrics between rendered and target audio.

Together, these components form a two-stage framework for synthesizer inversion as conditional generation over discrete synthesizer parameters. Experiments on Dexed show that, after supervised training, discrete diffusion substantially outperforms the continuous flow-matching baseline and is competitive with a carefully ordered autoregressive Transformer. An autoregressive order ablation confirms that the AR baseline is sensitive to heuristic parameter ordering, whereas discrete diffusion avoids this fixed-order design choice. GRPO fine-tuning with rendered-audio rewards then substantially improves out-of-domain matching on NSynth, reducing multiple audio-distance metrics and further improving CLAP distance when optimized with CLAP and CREPE rewards. These results demonstrate that our method is a strong and competitive approach for synthesizer inversion.

\section{Background}\label{sec:background}

\subsection{Synthesizer Inversion}

Synthesizer inversion, also known as parameter estimation or sound matching, was initially formulated as black-box optimization using genetic, evolutionary, or particle-swarm methods\cite{horner1993machine,garcia2001growing,mitchell2005frequency,heise2009automated}. Such iterative search often struggles to scale to modern high-dimensional synthesizers and may not transfer readily across parameter spaces.

\looseness=-1
Neural approaches instead predict synthesizer parameters directly from audio\cite{esling2019flow,le2021improving,chen2022sound2synth}. Hayes et al.\ further use approximately equivariant flow matching to jointly model synthesizer parameters and MIDI information such as pitch, velocity, and duration\cite{hayes2025audio}; we adopt this more general setting. It supports OOD audio with varying pitch and temporal length and therefore better reflects practical inputs from sources beyond the target synthesizer. However, a standard benchmark for this practical setting is still lacking.

A complementary direction uses differentiable synthesizers based on DDSP\cite{engel2020ddsp}, allowing audio-domain losses to be backpropagated to synthesis parameters\cite{masuda2023improving,uzrad2024diffmoog,yang2023white}. Han et al.\ similarly study parameter and perceptual audio losses for inverse sound matching\cite{han2024learning}. These methods typically require a differentiable white-box implementation, whereas our renderer is treated as a non-differentiable black box.

\subsection{Discrete Diffusion}

Discrete diffusion connects BERT-style masked language modeling and its extension to generation\cite{ghazvininejad2019mask,wang2019bert,nie2025large} with diffusion over discrete state spaces, particularly masked or absorbing-state processes\cite{austin2021structured,he2023diffusionbert}.

\looseness=-1
In music AI, discrete diffusion has been explored for symbolic music generation\cite{wang2025adaptive,yi2026vitex,plasser2023discrete}, and related masked frameworks have been used for biological sequence design\cite{wang2024dplm,gruver2023protein}. These domains involve constrained, strongly coupled, and structured discrete variables, resembling the dependencies among synthesizer parameters. This motivates masked discrete diffusion for synthesizer inversion, where it has not yet been systematically studied.

\begin{figure*}[t]
  \centering
  \includegraphics[alt={Training masks random synthesizer parameter tokens and predicts them from target audio and MIDI; inference starts fully masked and repeatedly selects a position by normalized entropy before sampling its parameter token},width=\textwidth]{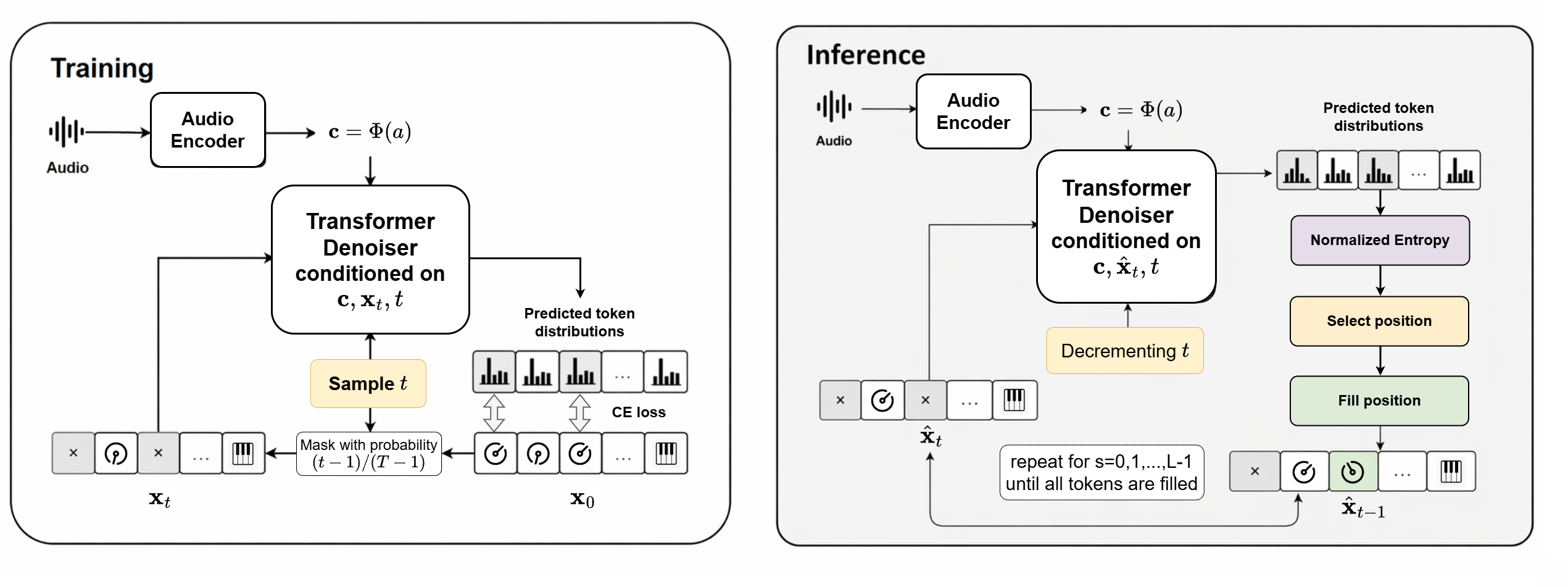}
  \caption{Training and inference of our supervised discrete diffusion model. Training randomly masks parameter tokens and applies supervision only to masked positions; inference starts from a fully masked sequence and iteratively fills the most confident position based on normalized entropy. Different icons represent parameter tokens and MIDI tokens.}
  \label{fig:method_overview}
\end{figure*}

\subsection{Reinforcement Learning for Music AI}

RL is useful when the desired objective is non-differentiable or poorly represented by supervised labels. GRPO obtains learning signals by comparing groups of candidates for the same input, avoiding an additional critic\cite{shao2024deepseekmath}. In music AI, RL has been applied to symbolic, real-time, and controllable music generation\cite{wang2025notagen,wu2025adaptive,cideron2024musicrl}, where musical objectives can be difficult to express through token-level losses.

\looseness=-1
For audio tasks, RL can directly optimize rewards computed from rendered audio through non-differentiable metrics such as perceptual similarity. SynthRL applies this idea to synthesizer inversion under fixed performance conditions\cite{shin2025synthrl}, but does not jointly model MIDI information and is therefore less suited to OOD audio with varying pitch and duration. Recent work also applies RL to masked discrete diffusion for language reasoning and unmasking-policy learning\cite{zhao2025d1,hong2026improving}; we instead use rendered-audio rewards for black-box synthesizer inversion while jointly predicting synthesizer and MIDI tokens.

\section{Method}\label{sec:method}

We use a two-stage framework for Dexed synthesizer inversion. First, we formulate inversion as conditional generation over discrete synthesizer and MIDI tokens, and train a masked discrete diffusion model with supervised token-level objectives. Second, we fine-tune the supervised model with GRPO using rewards computed from rendered audio.

\subsection{Tokenization}

We formulate Dexed inversion as conditional generation over discrete synthesizer tokens. Dexed\footnote{\url{https://asb2m10.github.io/dexed/}} is an open-source recreation of the Yamaha DX7 FM synthesizer and exposes 156 VST parameters. We represent a synthesizer patch as a token sequence in which each position corresponds to one synthesizer or MIDI control. In our schema, 102 Dexed controls are modeled and predicted as learnable discrete tokens, while the remaining 54 parameters are fixed to their default values during decoding and rendering. We also predict three MIDI-domain tokens: pitch, velocity, and duration. The target sequence is denoted as \( \bm{x}_0 = ( x_0^{(1)}, \ldots, x_0^{(L)} ) \), where each token \( x_0^{(i)} \) represents the value assigned to the \(i\)-th control. The target sequence has length \( L = 105 \).

Each token \( x_0^{(i)} \) takes values from its own vocabulary \( \mathcal{V}^{(i)} \). Under the frozen token schema, continuous-valued controls are discretized into bins; in our current Dexed setup, quantized controls use 25 bins. For supervised training, each clean token \( x_0^{(i)} \) is converted into a target distribution \( \tilde{Q}^{(i)} \) over \( \mathcal{V}^{(i)} \). In the case of purely categorical controls, \( \tilde{Q}^{(i)} \) is a one-hot distribution. In the case of MIDI tokens and numerically ordered quantized controls, \( \tilde{Q}^{(i)} \) is a locally smoothed target distribution to reflect the ordering of nearby values.

Given target audio \( \bm{a} \), we extract a normalized log-mel condition \( \bm{c}_\mathrm{audio} = \Phi(\bm{a}) \) and model \( p_\theta( \bm{x}_0 \mid \bm{c}_\mathrm{audio} ) \). We use a generative formulation because Dexed inversion is intrinsically one-to-many: different FM routings and parameter combinations can produce perceptually similar timbres.

\subsection{Discrete Diffusion Pretraining}

Our generative formulation of choice is masked discrete diffusion that models synthesizer parameters directly in token space. The model learns to recover masked parameter tokens from the audio condition and the currently visible tokens, rather than generating parameters in a fixed autoregressive order or through a continuous relaxation.

The total number of timesteps is \(T=L\). During training, we sample a timestep \(t\sim\mathrm{Uniform}(\{1,\ldots,T\})\) and set the masking probability \( p_\mathrm{mask}(t) = (t-1)/(T-1) \). We then sample a binary mask \( \bm{m} \sim \mathrm{Bernoulli}( p_\mathrm{mask}(t) )^L \) independently for each token position and construct the partially masked sequence \( \bm{x}_t \) as

\begin{equation}
\bm{x}_t = \bm{m} \odot \texttt{[MASK]} + (\bm{1} - \bm{m}) \odot \bm{x}_0
\end{equation}

The discrete diffusion model outputs a categorical distribution for every token position: \( p_\theta( \bm{x}_0 \mid \bm{c}_\mathrm{audio}, \bm{x}_t, t ) \) given the audio condition, the masked sequence, and the current timestep. Supervision is applied only to masked positions using the target distributions \( \tilde{\bm{Q}} \). The supervised objective is a discrete diffusion loss:

\begin{equation}
\mathcal{L}_{\mathrm{DD}}
=
\frac{
    \sum_i m^{(i)} w^{(i)}
    \operatorname{CE}( \tilde{Q}^{(i)}, \hat{Q}_\theta^{(i)} )
}{
    \sum_i m^{(i)} w^{(i)} + \epsilon_{\mathrm{DD}}
},
\label{eq:dd_loss}
\end{equation}
where
\begin{equation}
\hat{Q}_\theta^{(i)}
=
p_\theta( \bm{x}_0 \mid \bm{c}_\mathrm{audio}, \bm{x}_t, t )^{(i)}
\label{eq:dd_pred}
\end{equation}

Here \(w^{(i)} \in \mathbb{R}^+\) is a token loss weight used to emphasize parameters with stronger effects on the rendered audio.

Effectively, this design trains the model to recover arbitrary subsets of synthesizer parameters given the audio and the rest of parameters.

\subsection{Discrete Diffusion Decoding}

Inference starts from a fully masked sequence \( \hat{\bm{x}}_T = ( \mathrm{[MASK]}, ..., \mathrm{[MASK]}) \) and fills one token at each decoding step to obtain \( \hat{\bm{x}}_{t-1} \) from \( \hat{\bm{x}}_t \), for \(t=T,\ldots,1\). At each step \(t\), we obtain the posterior distribution:
\begin{equation}
\hat{\bm{Q}}_\theta = p_\theta( \bm{x}_0 \mid \bm{c}_\mathrm{audio}, \hat{\bm{x}}_t, t ).
\end{equation}
Since \( \hat{\bm{Q}}_\theta \) provides a posterior categorical distribution vector \( \hat{Q}_\theta^{(i)} \in \mathbb{R}^{|\mathcal{V}^{(i)}|} \) for each masked position \( i \), we choose a single masked position to sample and commit according to the normalized entropy confidence \( \gamma^{(i)} \) calculated from the Shannon entropy \( H^{(i)} \):
\begin{equation}
\begin{aligned}
H^{(i)} &= - ( \hat{Q}_\theta^{(i)} )^\top \log \hat{Q}_\theta^{(i)},\\
\gamma^{(i)} &= 1-\frac{H^{(i)}}{\log |\mathcal{V}^{(i)}|}.
\end{aligned}
\end{equation}
We fill the unresolved position with the largest \(\gamma^{(i)}\) using its most likely token to obtain \( \hat{\bm{x}}_{t-1} \) and repeat until all \(L\) tokens are resolved in \( \hat{\bm{x}}_0 \). When comparing unresolved positions, we use the \(\log |\mathcal{V}^{(i)}|\) normalization to make the confidence score aware of the vocabulary size.

\subsection{GRPO Fine-Tuning}

Token-level supervision does not directly optimize rendered audio quality, and Dexed is a non-differentiable black-box renderer. We therefore fine-tune the supervised discrete diffusion model with GRPO using rendered-audio rewards. In discrete diffusion decoding for GRPO, a full generation trajectory contains two types of decisions at each step: which unresolved parameter position to fill next, and which token value to assign at that position. The second decision is directly given by the predicted categorical distribution at the selected position, whereas the first is sampled stochastically from the normalized-entropy-based position distribution. To generate candidate trajectories for reward computation, we use stochastic best-of decoding: at each step, the position to fill is selected epsilon-greedily based on normalized-entropy confidence, and the token at that position is sampled from the predicted categorical distribution with top-\(k\) and minimum-probability filtering. This allows exploration of multiple plausible sequences, which are then rendered and evaluated with audio-domain metrics.

A trajectory \(\tau=(i_T,c_T,\ldots,i_1,c_1)\) records the sampled position and synthesizer parameter token choice at each decoding step. Its log-probability is accumulated from the stochastic position policy and token policy:
\begin{equation}
\begin{aligned}
\log\pi_\theta(\tau \mid \bm{c}_\mathrm{audio})
=
\sum_{t=1}^{T}
\left[
    \log \pi_\theta( i_t \mid \bm{c}_\mathrm{audio}, \hat{\bm{x}}_t, t )
    \right. \\ \left.
    +
    \log\pi_\theta( c_t \mid i_t, \bm{c}_\mathrm{audio}, \hat{\bm{x}}_t, t )
\right].
\end{aligned}
\end{equation}

For each audio target \( \bm{a} \), we sample \(K\) trajectories, decode and render them as \( \hat{\bm{a}}_k = \operatorname{Dexed}(\hat{\bm{x}}_{0,k}) \), and compute a reward from audio metrics. The distance terms $\mathcal{D}$ include wMFCC, CLAP embedding distance, CREPE embedding distance, multi-scale spectrogram distance, and spectral optimal transport distance, while RMS envelope cosine similarity is used as a similarity term:
\begin{equation}
\begin{aligned}
r_k
&=
-\sum_{j\in\mathcal{D}}\lambda_j d_j( \hat{\bm{a}}_k, \bm{a} )
\\
&\quad
+\lambda_{\mathrm{rms}}s_{\mathrm{rms}}( \hat{\bm{a}}_k, \bm{a} ).
\end{aligned}
\end{equation}
GRPO normalizes the rewards \( r_k \) for the same target audio using their mean \( \mu_{\bm{r}} \) and standard deviation \( \sigma_{\bm{r}} \):
\begin{equation}
A_k=\frac{r_k-\mu_{\bm{r}}}{\sigma_{\bm{r}} + \epsilon_{\mathrm{A}}},
\end{equation}
Because the reference log-probability is detached, it is constant with respect to \(\theta\) for sampled trajectories, and the implemented surrogate loss is equivalent up to an additive constant to
\begin{equation}
\mathcal{L}_{\mathrm{GRPO}}
=-\mathbb{E}_k\!\left[(A_k-\beta)\log\pi_\theta(\tau_k\mid\bm{c}_\mathrm{audio})\right].
\end{equation}
This GRPO-style objective combines relative-reward optimization with trajectory log-probability regularization and shifts optimization toward rendered-audio matching.

\section{Experiments}\label{sec:experiments}

\subsection{Datasets}

We divide the experimental data into in-domain and out-of-domain (OOD) subsets. For the in-domain data, we use the Dexed parameter-audio dataset curated by Le Vaillant and Dutoit for SPINVAE-2~\cite{le2024latent}, and follow their augmentation strategy by applying random perturbations to the synthesizer parameters and rendering each preset under four randomly sampled MIDI conditions. The Dexed portion contains 860,144 audio-parameter pairs, corresponding to approximately 215,036 preset groups. We split the data into training, validation, and test sets with an approximate 90\% / 5\% / 5\% ratio. During splitting, all MIDI renderings of the same preset are assigned to the same split rather than distributed across different splits, which avoids information leakage caused by shared underlying parameter presets.

For the OOD data, we use the NSynth dataset and directly adopt its official train/valid/test split. We treat these audio clips as external prompt audio without parameter annotations in order to evaluate the model's generalization to real OOD timbres. For GRPO fine-tuning, NSynth train audio serves as target prompts and NSynth valid audio is used for validation. No parameter annotations are used in this stage; rewards are computed only by comparing each target prompt with audio rendered by Dexed from the sampled parameters.

\subsection{Models}

We compare three conditional generative models for the same synthesizer inversion task: an autoregressive Transformer (AR), a discrete diffusion model (DD), and a continuous flow matching model (FM). All models take normalized log-mel spectrograms as audio conditions, and the audio-conditioning modules are trained jointly with the parameter generator rather than used as frozen pretrained feature extractors. AR and DD encode the input with a five-block 2D CNN, followed by a 12-layer Transformer encoder and a 12-layer parameter-token decoder (\(d_\mathrm{model}=512\), eight heads) with cross-attention to the encoded audio features.

\subsubsection{Autoregressive Model}

Remarkably, to the best of our knowledge, existing synthesizer inversion work has not systematically studied an autoregressive generation framework. SynthRL \cite{shin2025synthrl} uses a one-shot non-autoregressive Transformer architecture, where the model predicts all synthesizer parameters in parallel conditioned on the input audio. We build on its audio-conditioned encoder-decoder Transformer framework and reformulate parameter prediction as an autoregressive generation problem, so that the model generates parameter configurations token by token according to a fixed order.

The AR baseline uses the discrete token schema and appends an additional EOS token, resulting in a sequence length of 106. During training, the target token sequence is shifted to the right and prepended with a BOS token, and the decoder predicts the next token under a causal mask. During inference, the model starts from the BOS token and an empty history, and greedily selects the most likely token at each position.

The training objective is token-level cross entropy. For numerical parameters and MIDI tokens, we apply Gaussian smoothing convolution to convert one-hot targets into locally smoothed target distributions, making the learning target easier to optimize.

\subsubsection{Discrete Diffusion Model}

In the experiments, the discrete diffusion model (DD) uses the same type of audio-conditioned encoder-decoder Transformer framework and audio encoder as the AR baseline. Unlike the AR decoder, the DD decoder does not use a causal mask, so each position can attend to the currently visible tokens at all other positions. To represent the diffusion step, the model learns a time embedding for each \(t\) and adds it to the token embeddings.

Training and inference follow the masking and confidence-based decoding procedures in \secref{sec:method}. We use the same smoothing strategy as the AR baseline for numerical parameters and MIDI tokens.

\subsubsection{Flow Matching Model}

The FM baseline follows the approximately equivariant flow matching framework of Hayes et al. \cite{hayes2025audio}, using its AST-style spectrogram encoder and ApproxEquivTransformer vector field (both 12 layers, \(d_\mathrm{model}=512\), and eight heads). Unlike AR and DD, which operate directly in discrete token space, FM formulates synthesizer inversion as conditional generation in a continuous space. We adapt its target space to our Dexed/MIDI setting: numerical Dexed controls remain continuous scalars, categorical controls are expanded into one-hot continuous blocks, and MIDI-related information is incorporated into the same denoise space as continuous variables.

\subsection{Evaluation Metrics}

\looseness=-1
For evaluation, we follow several audio similarity metrics used by Hayes et al. \cite{hayes2025audio} to measure differences between the target audio and the audio rendered from predicted parameters. wMFCC computes MFCC features and aligns them with dynamic time warping; multi-scale spectrogram distance (MSS) compares mel spectrograms across multiple time-frequency resolutions; spectral optimal transport (SOT) normalizes each spectral frame as an energy distribution and compares spectral energy distributions using an optimal-transport distance. We also report MFCC distance and RMS envelope cosine similarity. In addition, following Tian et al. \cite{tian2025assessing} on perceptual audio similarity assessment, we use CLAP embedding cosine distance as a higher-level audio similarity metric. Except for RMS envelope cosine similarity, lower values indicate better matching for all distance-based metrics.

\begin{table*}[!t]
  \centering
  \small
  \renewcommand{\arraystretch}{1.12}
  \resizebox{\textwidth}{!}{%
  \begin{tabular}{lrrrrrrr}
    \hline
    \multicolumn{8}{c}{\textbf{In-Domain Dexed}} \\
    \hline
    Method & wMFCC \(\downarrow\) & MFCC13 \(\downarrow\) & MFCC40 \(\downarrow\) & MSS \(\downarrow\) & SOT \(\downarrow\) & RMS \(\uparrow\) & CLAP \(\downarrow\) \\
    \hline
    Autoregressive & \textbf{6.30} & \textbf{9.44} & \textbf{6.35} & \textbf{3.41} & \textbf{0.028} & 0.964 & \textbf{0.128} \\
    Flow Matching & 7.83 & 12.69 & 8.57 & 4.66 & 0.054 & 0.941 & 0.203 \\
    Discrete Diffusion (Stage 1) & 6.65 & 10.28 & 6.71 & 3.70 & 0.036 & \textbf{0.968} & 0.139 \\
    \hline
    DD-GRPO (Multi-Reward) & 10.92 & 17.45 & 9.30 & \textbf{4.48} & \textbf{0.055} & 0.942 & 0.241 \\
    DD-GRPO (CLAP+CREPE) & \textbf{10.57} & \textbf{17.08} & \textbf{9.23} & 4.73 & 0.057 & \textbf{0.943} & \textbf{0.228} \\
    \hline
    \multicolumn{8}{c}{\textbf{OOD NSynth}} \\
    \hline
    Method & wMFCC \(\downarrow\) & MFCC13 \(\downarrow\) & MFCC40 \(\downarrow\) & MSS \(\downarrow\) & SOT \(\downarrow\) & RMS \(\uparrow\) & CLAP \(\downarrow\) \\
    \hline
    Autoregressive & 10.85 & 17.94 & 9.61 & \textbf{4.15} & \textbf{0.066} & \textbf{0.916} & \textbf{0.439} \\
    Flow Matching & 12.60 & 21.41 & 12.21 & 5.33 & 0.086 & 0.870 & 0.523 \\
    Discrete Diffusion (Stage 1) & \textbf{10.04} & \textbf{17.14} & \textbf{9.36} & 4.41 & 0.071 & 0.898 & 0.452 \\
    \hline
    DD-GRPO (Multi-Reward) & \textbf{5.96} & \textbf{9.58} & \textbf{6.13} & \textbf{3.11} & \textbf{0.038} & \textbf{0.944} & 0.413 \\
    DD-GRPO (CLAP+CREPE) & 7.28 & 11.88 & 7.04 & 3.64 & 0.056 & 0.931 & \textbf{0.357} \\
    \hline
  \end{tabular}%
  }
  \caption{In-domain and OOD test performance. Metrics are averaged over 200 rendered examples for each split.}
  \label{tab:results}
\end{table*}

\subsection{Autoregressive Order Ablation}

Before the main model comparison, we conduct an ablation on the AR order. We compare our manually designed Dexed parameter order with randomly shuffled orders. From figure 3, the manually designed order consistently converges faster and achieves lower validation errors than the random orders, demonstrating that our proposed order is effective. At the same time, the results also show that the autoregressive baseline is sensitive to the generation order, suggesting that obtaining a strong AR baseline requires heuristic design and empirical validation.

\subsection{Training}

\begin{figure}[t]
  \centering
  \includegraphics[alt={Autoregressive order ablation curves comparing manual and random parameter orders},width=\linewidth]{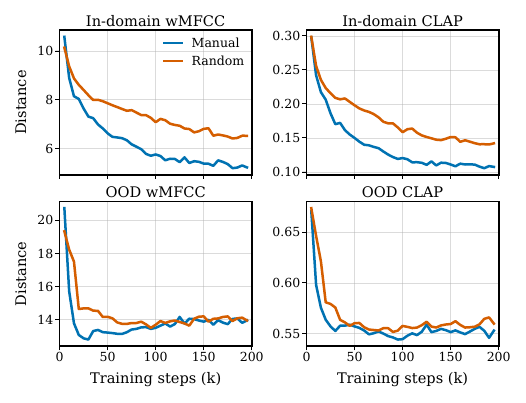}
  \caption{Validation curves for the AR order ablation. Each curve averages four seeds after aligning validation checkpoints to a shared training-step grid.}
  \label{fig:ar_order_ablation}
\end{figure}

We train AR, DD, and FM on the in-domain Dexed training set and evaluate them on both the in-domain validation split and the NSynth OOD validation set. For stages that require audio rendering, we use DawDreamer \cite{braun2021dawdreamer} to host the Dexed VST and batch-render audio from predicted synthesizer parameters and MIDI conditions. In the second stage, we initialize from the first-stage DD checkpoint and fine-tune it with GRPO using rendered-audio rewards. We use \(K=8\) candidates per prompt and \(\beta=0.1\); the first phase uses the multi-metric reward above, while the second keeps only CLAP and CREPE.

\subsection{Results}\label{sec:results}

\looseness=-1
Table~\ref{tab:results} reports the results after first-stage supervised training and second-stage GRPO fine-tuning. In the first-stage supervised setting, Flow Matching performs substantially worse than both the autoregressive model and the discrete diffusion model, supporting our hypothesis that encoding inherently discrete synthesizer parameters as logits in a continuous space introduces additional continuous relaxation and projection difficulties. On the in-domain Dexed test set, the autoregressive model slightly outperforms the first-stage discrete diffusion model on most metrics, whereas on the OOD NSynth dataset, discrete diffusion outperforms the autoregressive model on the MFCC-related metrics while the autoregressive model remains better on MSS, SOT, RMS, and CLAP.

For synthesizer inversion, OOD evaluation is more important than in-domain evaluation. Although our split avoids direct information leakage, higher-level similarity is still difficult to eliminate from the in-domain data. In practical use cases, the input audio is usually not rendered by the target synthesizer itself, but may instead come from other instruments, other synthesizers, sample libraries, or real recordings. Therefore, the OOD setting is closer to real-world usage.

Second-stage GRPO fine-tuning demonstrates the effectiveness of audio-domain rewards. On the OOD dataset, GRPO with the multi-reward objective substantially reduces several audio distance metrics, showing that GRPO can use audio feedback from the black-box synthesizer to shift the optimization objective from pure parameter-token matching toward direct audio similarity. Because GRPO is trained on NSynth prompts, this shift moves the policy away from the in-domain Dexed preset distribution, explaining the degradation of in-domain metrics. When we continue training in the second phase using only CLAP and CREPE rewards, the CLAP distance further decreases.

\section{Discussion}\label{sec:discussion}

Although the autoregressive model can achieve performance close to discrete diffusion on the relatively fixed Dexed synthesizer setting, our ablation experiments show that its performance is highly sensitive to the generation order of parameter tokens. This heuristic design may not transfer reliably to more complex synthesizers, where the number of parameters increases, the module structure becomes more complex, and parameter dependencies become more irregular. In contrast, discrete diffusion avoids the fixed-order assumption at the modeling level and handles synthesizer parameters in a more symmetric and flexible way.

We also observe that allowing the first-stage supervised model to overfit slightly does not necessarily produce the best objective audio metrics, but can lead to better subjective listening quality. One possible explanation is that human-designed synthesizer preset datasets contain not only mappings from parameters to audio, but also the aesthetic preferences of sound designers. The prior parameter distribution learned by the supervised model is therefore shaped by human sound design practices rather than being uniform. As a result, following this data distribution more closely may produce sounds that better match user expectations in real creative scenarios, even if some low-level audio distance metrics are not optimal.

This observation suggests that human-designed preset datasets may be more suitable than randomly sampled parameter datasets for synthesizer inversion. For music-creation-oriented synthesizer inversion, this implicit aesthetic prior may be as important as audio similarity itself.

\section{Conclusion}\label{sec:conclusion}

In this paper, we study discrete diffusion modeling for audio synthesizer inversion and introduce GRPO fine-tuning with audio-domain rewards from a black-box synthesizer renderer. Experimental results show that discrete diffusion substantially outperforms Flow Matching while avoiding the need to predefine a parameter generation order. It also achieves performance close to the autoregressive model and performs better on some metrics in the OOD setting. GRPO fine-tuning further improves OOD audio matching metrics, demonstrating the effectiveness of audio-domain rewards for non-differentiable synthesizer inversion.

\section{AI Usage Statement}

ChatGPT was used to assist with implementing the training code and writing and revising this paper.

\section{Acknowledgements}

I thank Ziyuan Zhao and Lejun Min for insightful technical discussions, Liwei Lin for assistance with the paper's figures, and my advisor, Mark Nicholas Grimshaw-Aagaard, for his continued support and intellectual guidance.


\bibliography{ISMIRtemplate}

\end{document}